\documentclass[aps,superscriptaddress,nofootinbib]{revtex4}
\usepackage{hyperref}
\usepackage{amsmath}
\usepackage{graphicx}
\usepackage[normalem]{ulem}
\usepackage{url}
\usepackage{color}

\usepackage[charter]{mathdesign}
\DeclareFontFamily{U}{dutchcal}{\skewchar\font=45 }
\DeclareFontShape{U}{dutchcal}{m}{n}{<-> s*[1.0] dutchcal-r}{}
\DeclareFontShape{U}{dutchcal}{b}{n}{<-> s*[1.0] dutchcal-b}{}

\begin{document}
\title{Statistical treatment of nuclear clusters in the continuum}

\author{S. Mallik}
\email{swagato@vecc.gov.in}
\affiliation{LPC Caen IN2P3-CNRS/EnsiCaen et Universite, Caen, France}
\affiliation{Physics Group, Variable Energy Cyclotron Centre, 1/AF Bidhan Nagar, Kolkata 700064, India}
\author{F. Gulminelli}
\affiliation{LPC Caen IN2P3-CNRS/EnsiCaen et Universite, Caen, France}

\begin{abstract}
 The evaluation of the sub-saturation nuclear equation of state at finite temperature  requires a proper state counting of the internal partition sum of nuclei which are immersed in the background of their continuum states. This classical statistical problem is addressed within the self-consistent mean-field approximation, which naturally accounts for isospin and effective mass effects in the nuclear density of states. The nuclear free energy is decomposed into bulk and surface terms, allowing a simple analytical prescription for the subtraction of gas states from the nuclear partition sum, that avoids double counting of unbound single particle states.
We show that this correction leads to a sizeable effect in the composition of matter at high temperature and low proton fractions, such as it is formed in supernova collapse, early proto-neutron star evolution, as well as laboratory experiments. Specifically, the energy stored in the internal nuclear degrees of freedom is reduced, as well as the mass fraction of heavy clusters in the statistical equilibrium. The gas subtraction prescription is compared to different phenomenological methods proposed in the literature, based on a high energy truncation of the partition sum. We show that none of these methods satisfactorily reproduces the gas subtracted level density, if the temperature overcomes $\approx 4$ MeV.
\end{abstract}

\maketitle

\section{Introduction}

A correct statistical mechanics description of nuclear clusters is very important in the study of the nuclear equation of state
below saturation density \cite{Burgio2018}. Dilute and warm matter is dominated by the presence of bound nuclei which
are never negligible in a very large domain of densities $n_B\approx 10^{-6}-10^{-1}$ fm$^{-3}$, temperatures $k_B T\approx 0-10$ MeV, and proton fractions $y_p\approx 0.01-0.6$ \cite{Raduta2010,Hempel2010,Furusawa2011}, and prevail over free nucleons in most conditions encountered in nature \cite{Hempel2012}. These latter encompass pre-supernova collapse, post-bounce dynamics, proto-neutron star evolution and neutron star mergers\cite{Fischer2014}, as well as multi-fragmentation reactions produced in the laboratory by heavy ion collisions \cite{Borderie2019,Dasgupta_book}.

In all these situations, nuclei of all sizes are believed to coexist in thermal and chemical equilibrium with a nuclear gas of protons and neutrons in strong and electromagnetic equilibrium . In the astrophysical context, the simultaneous presence of nuclei and nucleons in the same physical volume is due to the gravitational pressure of the star. In the case of heavy ion collisions, a finite effective volume appears in the statistical treatment due to the freeze-out concept in an expansion dynamics: this volume physically measures the spatial extension of the dynamical system at the earliest time when the strong interactions can be neglected and the nuclear abundancies are consequently frozen.

In both scenarii, the theoretical estimation of the relative abundances of nuclei and nucleons depends both on the thermodynamic condition, and on the treatment of the internal partition sum of the nuclei.
Indeed, in all the present models of the sub-saturation equation of state, a fully microscopic treatment of baryonic matter is out of scope, and nuclei and nucleons are considered as separate degrees of freedom in the statistical approach.  This naturally opens the question of interactions and more generally statistical correlations between the two components represented by nuclei and nucleons. The standard view of the problem is that the nuclear interaction is exhausted by the formation of bound states of one side, and by the existence of a non-zero density and momentum-dependent mean-field modifying the energetics of the free nucleons on the other side. Residual interactions between nuclei and nucleons are additionally included in the excluded volume approximation\cite{Raduta2010,Hempel2010,Furusawa2011}.

However, the independence of the two components is at least violated by the fermionic nature of nucleons. In a simplified mean-field picture of nuclear structure, single particle states occupied by free nucleons cannot be simutaneously accessed by nucleons bound in clusters, which are promoted to higher energy states, leading to a decrease of the cluster binding.  An important theoretical \cite{Typel2010,Ropke2011,Pais2019} as well as experimental \cite{Qin2012,Hempel2015,Bougault2020,Pais2020} effort was devoted in the recent literature to the evaluation of the shift in the light nuclei ground state binding energy  induced by this Pauli blocking effect, and it was shown that it leads to an important suppression of light cluster abundancies at the so called Mott density, with potential important consequences on the neutrino dynamics in different astrophysical environnements \cite{Fischer2014,Arcones2008,Furusawa2017NPA}.

No sizeable  ground state binding energy shift is expected in heavy clusters, because of the negligible overlap between the ground state of heavy nuclei and the single particle states of the nuclear gas.   However, this is not true for the excited nuclear states beyond nucleon separation, which are abundantly populated at finite temperature.
In the statistical treatment,  these states are accounted for in the internal nuclear partition sum, and correspond to an excitation energy allowing the emission of one or several nucleons in the continuum. To avoid  overcounting of continuum configurations, a certain fraction of the compound nucleus state density at high excitation should therefore not be counted in the nuclear partition sum.

This double counting problem of the continuum states was noticed very early in the nuclear astrophysics literature \cite{Bethe1978,Fowler1978}, and a thermodynamically consistent solution was  proposed already in the late seventies
by Tubbs and Koonin \cite{Koonin1979} and exploited with  microscopic self-consistent methods by Bonche, Levit and Vautherin \cite{Bonche1984}.

In the seminal work by Tubbs and Koonin \cite{Koonin1979} it was shown that the continuum subtraction procedure only weakly affects the partition sum and the average excitation energy of $^{56}Ni$ in thermodynamic conditions corresponding to early pre-supernova dynamics. Probably for this reason, in most of the subsequent astrophysical equation of state literature, corrections have been applied only to the surface free energy, and
 unmodified bulk partition sums have been employed \cite{LS,Shen,Shen2011,Furusawa2011,Furusawa2017}, with very few exceptions \cite{Nandi2010}.
 A phenomenological prescription was adopted in some works, consisting in cutting the internal cluster partition sum to the nucleon separation energy \cite{Gulminelli2015} or to the total binding energy of the nucleus \cite{Hempel2010,Hempel2012}.

However, most of the supernova and NS mergers dynamics corresponds to thermal conditions different from the ones of the early calculations, and it was recently shown \cite{Raduta2019} that the treatment of continuum states in the internal cluster partition sum is at the origin of the most important model dependence observed in sub-saturation equations of state \cite{Buyu2013}.

The importance of the issue was also recognized  in the multifragmentation literature \cite{Borderie2019}.
A simpler prescription than the continuum subtraction technique by Bonche et al. was introduced  by Randrup and Koonin in the early years of multifragmentation research \cite{Randrup1987a,Randrup1987b}, and shown to be necessary for a coherent interpretation of the whole set of fragmentation observables \cite{Piantelli2005,Borderie2013}. This prescription consists in introducing a modified internal level density function for the fragments as:

\begin{equation}
\rho(A,E^*)=\rho_{FG}(A,E^*) e^{-E^*/T_{lim}} , \label{eq:tlim}
\end{equation}

where $\rho_{FG}$ is the standard level density of a free Fermi gas in the Sommerfeld approximation for a nucleus of mass number $A$ and excitation energy $E^*$ (see below for precise definitions), and $T_{lim}$ is a limiting temperature, which in principle can be calculated consistently in a given microscopic model as the finite system analog of the thermodynamical transition temperature \cite{Baldo2004}, but in practice it is usually determined from a fit of the experimental data.

In this paper, we work out  explicitly the internal cluster partition sum in the self-consistent mean field approximation applying  the continuum subtraction. We  first calculate the level density  by inverse Laplace transform of the partition sum, and show  that this expression considerably deviates from the Fermi gas expression even in the bulk limit, due to mean field and effective mass effects. In particular, accounting for the effective mass considerably reduces the level density even at low temperature, and consequently reduces the amount of excitation energy $E^*$ stored in the clusters at the temperatures  relevant for both multifragmentation and supernova evolution.

As a second step, we  study the impact of the continuum subtraction on the $E^*(T)$ relation and on the mass fraction of the clusters in different thermodynamic conditions. We will see that the different empirical prescriptions employed in the literature are not equivalent to the consistent gas subtraction. Analytical expressions for the bulk free energy are given, that can be easily implemented in realistic equations of state.

 \section{Calculation in the Wigner-Seitz cell} \label{sec:WScell}
We first consider the simple case of a unique ion confined at finite temperature $T$ in a cell of finite volume $V$, the background of its continuum states extending out of the spatial extension of the nucleus. The total Helmholtz free energy of the cell is decomposed as the sum of the nucleus or fully ionized ion, the nucleon gas, and the electrons (if we consider stellar matter, that is neutralized by a uniform electron background):

\begin{equation}
F_{cell}=F_N+F_g+F_e . \label{eq:fcell}
\end{equation}

The nuclear interaction between the nucleus and the nucleon gas, and the Coulomb interaction between the protons and the electrons, are all included in the term $F_N$, such that the free nucleon and electron component are simply given by
\footnote{We denote with capital letters the (free) energy per ion, e.g.
F, small letters e.g. $f$ indicates quantities per baryon, while the notation $\mathcal F$ is used for the free energy density.}
:

\begin{equation}
 F_ g= V \mathcal F_g \; ; \;  F_e = V \mathcal F_e \ . \label{eq:fgas}
\end{equation}

In this equation, $  \mathcal F_{e}$ is the free energy density of a uniform electron gas at density $n_e$ \cite{CoxGiuli},
and $  \mathcal F_{g}$ corresponds to a uniform nucleon  gas at
 proton (neutron) density $n_{g,p(n)}$, to be specified later.

In the presence of continuum states, the evaluation of the ion contribution $F_N$ has to be handled with care, because a simple state counting in the ion field includes contributions which are already accounted in the gas component $F_g$. The ion and the gas correspond to very different densities, but are associated to the same chemical potentials. For this reason the gas subtraction is easier to calculate in the grand canonical ensemble. The grand canonical thermodynamic potential $\Omega_N$ is linked to the Helmotz free energy $F_N$ by the following Legendre transform:

\begin{equation}
 F_ N= \Omega_N +\mu_n N_n + \mu_p N_p  \; ; \;    \Omega_N=- T \ln {\cal{Z}}^{(N)}_{\beta\mu_n\mu_p}\ ,
\end{equation}

where $\beta=T^{-1}$, $N_n$ and $N_p$ are the neutron and proton numbers of the nucleus, and $\mu_{n(p)}$ is the neutron (proton) chemical potential. For a given set  of intensive parameters $\{ \beta,\mu_n,\mu_p\}$, two different solutions exist in the finite size system, namely a purely gazeus homogeneous solution $\Omega_g$, and the full nucleus-plus-gas solution $\Omega_{Ng}$.  If those solutions are obtained from some microscopic variational theory  \cite{Koonin1979,Bonche1984,Suraud1987,De2001,Sil2002,Nandi2010} and additivity is assumed as in Eq.(\ref{eq:fcell}) \footnote{Additivity is justified by the fact that the possible gas-nucleus interaction are accounted for in the thermodynamical potential of the cluster.}, the thermodynamical potential of the bound part of the nucleus can be calculated by simple subtraction of the two:

\begin{equation}
  \Omega_N= \Omega_{Ng}-\Omega_g \ , \label{eq:subtraction}
\end{equation}

The calculation of the partition sum was performed in the semiclassical limit with an external spherical well potential in ref. \cite{Koonin1979}, and in successive works it was obtained with the self-consistent Hartree-Fock \cite{Bonche1984} and Thomas-Fermi   \cite{Suraud1987,De2001,Sil2002,Nandi2010} theory using Skyrme forces. To have a simple analytic expression of the nuclear partition sum, and be able to extend the calculation to account for full nuclear distributions, we work it out in the next subsection in the compressible liquid drop model.

\subsection{Analytical gas subtraction in the compressible liquid drop formalism}

If the total free energy $F_{cell}$ is known through some microscopic ab-initio modeling, the in-medium modified nucleus free energy $F_N$ is simply obtained as the excess free energy of the cell, with respect to the dominant contribution given by an homogeneous electron and nucleon gas. However in the general case $F_{cell}$ is not known, and a model for $F_N$ is in order. We here restrict to heavy and medium-heavy nuclei, for which a so called  “leptodermous” \cite{brack} expansion can be safely applied, and the interactions between the bound and unbound single particle states can be treated as a surface term \cite{papa,bira}.
The decomposition of the nuclear free energy reads \footnote{The convention $c=k_B=1$ for the speed of light and the Boltzmann constant  is used throughout the paper.} :

\begin{equation}
F_N= F_N^{id}+F_N^{bulk}+F_N^{coul}+ F_N^{surf} . \label{eq:leptodermous}
\end{equation}

Here, $F_N^{id}$ is the center-of-mass translational contribution \cite{LS} :
\begin{equation}
F_N^{id}= T \left ( \ln \frac{\lambda_N^3}{V g_ N} - 1 \right ) \label{eq:fideal}
\end{equation}
where $g_N=2J_N+1$ is the ground state degeneracy, and $\lambda_N=\hbar(2\pi/(M_nT))^{1/2}$ is the de Broglie wavelength, with $M_N=N_n m_n + N_p m_p$ the bare bound ion mass.
The nuclear energy is decomposed into a bulk $F_N^{bulk}$ and a surface $F_N^{surf}$ part, possibly dependent on the external gas. In this formalism, the nucleus is considered as a portion of bulk nuclear matter at density $n_c=n_{c,n}+n_{c,p}$ and  isospin asymmetry  $\delta_c= (n_{c,n}-n_{c,p})/n_c$ occupying a finite spatial volume $V_N=A_N/n_c$, and finite size corrections are included in the interface part  $F_N^{surf}$ \cite{papa}.

Finally, $F_N^{coul}=E_N^{coul}$ is the temperature independent Coulomb energy:
\begin{equation}
F_N^{coul}=\frac{3}{5}\frac{e^2N_p^2}{4\pi\epsilon_0}(1-f_{WS})\left (\frac{4\pi}{3V_N}\right )^{1/3} ,
\end{equation}
where the electron screening factor $f_{WS}$ is obtained in the
  Wigner-Seitz approximation as:
\begin{equation}
  f_{WS}(\delta_c, n_e)=\frac 32\left ( \frac{2 n_e}{(1-\delta_c)n_c}\right )^{1/3} -
\frac 12\left ( \frac{2 n_e}{(1-\delta_c)n_c}\right ). \label{eq:screening}
\end{equation}

The total particle number $A_N$ comprises both bound and unbound nucleons according to:
\begin{equation}
A_N=N_n+N_p+n_g V_N , \label{eq:a_N}
\end{equation}
and similarly $Z_N=N_p+n_{g,p} V_N$, with the unbound nucleons density  $n_g=n_{g,n}+n_{g,p}$.
Particularizing  Eq.(\ref{eq:subtraction}) to the decomposition (\ref{eq:leptodermous}), the gas subtraction affects only the bulk part of the cluster partition sum. We express the bulk free energy density ${\mathcal F}_N^{bulk}=F_N^{bulk}/V_N$ in terms of the grand canonical thermodynamic potential $\omega_N=-T \ln z^N_{\beta,\mu_n,\mu_p}$, as:

\begin{equation}
{\mathcal F}_N^{bulk}=\omega_N+\mu_n \left (n_{c,n}-n_{g,n}\right )+\mu_p \left (n_{c,p}-n_{g,p}\right ) \ , \label{eq:fbulk}
\end{equation}

where we still have to specify $\omega_N$. To this aim, we follow the standard derivation of the self-consistent mean field theory \cite{Vautherin1996}. For independent fermions, the partition sum is factorized as:
\begin{equation}
z^0_{\beta,\mu_n,\mu_p}=\prod_{q=n,p}\prod_{k}\bigg{\{}1+\exp\big{\{}\beta (\mu_q- e_{q,k})\big{\}}\bigg{\}} .
\end{equation}

Here, the second product runs over single particle states, and for bulk matter the single particle energies are given by
\begin{equation}
e_{q,k}=\frac{\hbar^2 k^2}{2m^*_{q}}+ U_{q} ,
\end{equation}

where the effective mass $m^*_q(n_n,n_p)$ and the mean-field potential $U_q(n_n,n_p)$ are calculated at the
densities satisfying the coupled self-consistent equations,

\begin{eqnarray}
 n_q &=& \sum_{k}\bigg{\{}1+\exp\big{\{}\beta (\mu_q- e_{q,k})\big{\}}\bigg{\}}^{-1} . \label{eq:nq}
\end{eqnarray}

In the thermodynamic conditions where matter is clusterized, these equations admit two solutions
$(n_{i,n},n_{i,p})$, with $i=c,g$,  corresponding to the two different phases that can coexist in infinite uncharged nuclear matter at equilibrium. They define two different partition sums $z^{0,c}_{\beta,\mu_n,\mu_p}$ and $z^{0,g}_{\beta,\mu_n,\mu_p}$.
In a finite system, according to the decomposition Eq.(\ref{eq:leptodermous}), $z^{0,c}_{\beta,\mu_n,\mu_p}$
corresponds to the bulk part of the cluster, which contains bound, resonant, and continuum states.
Because of the independence of the single-particle states, in order to sort out the continuum contribution,  we can simply write  $ z^{0,c}_{\beta,\mu_n,\mu_p}=z^{0,N}_{\beta,\mu_n,\mu_p}z^{0,g}_{\beta,\mu_n,\mu_p}$, where $z^{0,N}_{\beta,\mu_n,\mu_p}$ is the nucleus partition sum we are interested in.
Explicitating the sum over $k$ we have:

\begin{eqnarray}
\ln z^{0,N}_{\beta,\mu_n,\mu_p} &=&
 \frac{2}{h^3}\sum_{q=n,p}
\int d^3p \left (  \ln \bigg{[}1+\exp \beta\bigg{\{}\mu_q-\frac{p^2}{2m^*_{c,q}}-U_{c,q}\bigg{\}}\bigg{]} \right. \nonumber
\\
&-&  \left.  \ln \bigg{[}1+\exp \beta\bigg{\{}\mu_q-\frac{p^2}{2m^*_{g,q}}-U_{g,q}\bigg{\}}\bigg{]} \right ) .
\end{eqnarray}

Elementary manipulations lead to:

\begin{eqnarray}
\ln z^{0,N}_{\beta,\mu_n,\mu_p}
=\frac{2}{3}\beta \big{\{}\big{(}\xi_{c,n}+\xi_{c,p}\big{)}-\big{(}\xi_{g,n}+\xi_{g,p}\big{)}\big{\}} \ , \label{eq:z0}
\end{eqnarray}

where $\xi_{i,q}$ is the kinetic energy density of component $q$ in  phase $i$:

\begin{eqnarray}
\xi_{i,q}&=&  \sum_{k} \frac{\hbar^2k^2}{2 m^*_{i,q}}\bigg{\{}1+\exp\big{\{}\beta (\mu_q- e_{i,q,k})\big{\}}\bigg{\}}^{-1} \nonumber \\
&=&\frac{1}{2\pi^2} \bigg{(}\frac{2 m^*_{i,q}}{\beta \hbar^2}\bigg{)}^{5/2}F_{3/2}(\eta_{i,q}) ,
\end{eqnarray}
and
$\eta_{i,q}=\beta \left (\mu_q-U_{i,q}\right )$ is the corresponding effective chemical potential,
with $U_{i,q}=U_q(n_{i,n}, n_{i,p})$ the mean field which can depend on the particle densities $n_{i,q}$:

\begin{equation}
n_{i,q}=\frac{1}{2\pi^2} \Big{(}\frac{2 m^*_{i,q}}{\beta \hbar^2}\Big{)}^{3/2}F_{1/2}(\eta_{i,q}) , \label{eq:densities}
\end{equation}

and $F_{3/2}, F_{1/2}$ are Fermi functions.
Up to this point the derivation closely follows the one by Tubbs and Koonin \cite{Koonin1979} based on the independent particle model, where those authors took $U_{c,q}=U_0$, $U_{g,q}=0$ and $m^*_{i,q}=m$.

The use of a self-consistent mean field leads to two important modifications with respect to the seminal Tubbs and Koonin paper.
First, the nucleon gas is interacting, $U_{g,q}\ne 0$, and the effective masses are density dependent. This, as we will see in the next section, has an important effect on the partition sum.
 Another difference from the work of Ref.\cite{Koonin1979} arises in the computation of the free energies. Indeed, the mean field and independent particle formalisms correspond to the same state counting but to different expressions for the energy density, due to the rearrangement terms arising from the self-consistency of the mean-field.  This is most easily seen in the variational derivation of the finite temperature mean field theory \cite{Vautherin1996}, that we briefly recall.
Both the independent particle and the mean field partition sum can be obtained from a maxEnt principle, where the entropy density $\ln {\mathcal W}^{IPM}$ associated to independent single particle states is maximized under the constraint of given expectation values for particle number(s) and energy. In the independent particle case we have:

\begin{equation}
\ln z^{0}_{\beta,\mu_n,\mu_p}  = \ln {\mathcal W}^{IPM}- \beta \left (\epsilon^0-
\sum_{q=n,p}\mu_q n_{q}\right ) . \label{eq:IPM}
\end{equation}

Here, the single particle total energy density $\epsilon^0=\xi_{n}+\xi_{p}+U_{n}n_{n}+U_{p}n_{p}$, and
$U_{q}=U_q(n_{n},n_{p})$ is the mean field acting on particle type $q$ when the density is $n=n_{n}+n_{p}$ and the isospin asymmetry is $\delta= (n_{n}-n_{p})/n$.
Similarly for the mean field we can write:

\begin{equation}
\ln z^{mf}_{\beta,\mu_n,\mu_p}  = \ln {\mathcal W}^{IPM}- \beta \left (\epsilon^{mf}-
\sum_{q=n,p}\mu_q n_{q}\right ) \, \label{eq:MF}
\end{equation}

where $\epsilon^{mf}=\xi_{n}+\xi_{p}+v(n,\delta)$ is the mean field energy density at the same particle densities values. Because of rearrangement terms, in general $v\ne \sum_q U_q n_q$. Comparing Eqs.(\ref{eq:IPM}),(\ref{eq:MF}) we obtain the well-known expression for the grand canonical thermodynamical potential in the mean-field approximation \cite{Vautherin1996}:

\begin{equation}
\ln z^{mf}_{\beta,\mu_n,\mu_p} = \ln z^{0}_{\beta,\mu_n,\mu_p}- \beta \big{[}v(n,\delta)-U_n n_{n}-U_p n_{p}\big{]} . \label{eq:zmf}
\end{equation}

Applying the same gas subtraction as in Eq.(\ref{eq:z0}) we immediately get:

\begin{eqnarray}
\ln z^{mf,N}_{\beta,\mu_n,\mu_p}
&=&\frac{2}{3}\beta\big{[}\big{(}\xi_{c,n}+\xi_{c,p}\big{)}-\big{(}\xi_{g,n}+\xi_{g,p}\big{)}\big{]} \label{eq:gas_subtraction}
 \\
&-&\beta\big{[}v(n_c,\delta_c)-U_{c,n}n_{c,n}-U_{c,p}n_{c,p}\big{]} \nonumber \\
&+&\beta\big{[}v(n_g,\delta_g)-U_{g,n}n_{g,n}-U_{g,p}n_{g,p}\big{]} . \nonumber
\end{eqnarray}

Once the internal partition sum of the clusters is defined, all observables can be computed from general thermodynamical relations. Recalling that the particle numbers associated to the bound and resonant part of the clusters are  $N_q=(n_{c,p}-n_{g,p}) V_N$, we have for the bulk part of the Helmotz free energy:
\begin{eqnarray}
F_N^{bulk}(N_n,N_p)
&=&-T V_N\ln z^{mf,N}_{\beta,\mu_n,\mu_p}  +\mu_n N_n
+\mu_p N_p \label{eq:FN}\\
&=&V_N \bigg{ [}v(n_c,\delta_c)-v(n_g,\delta_g) - \sum_q \left ( U_{c,q}n_{c,q}-U_{g,q}n_{g,q}\right ) \bigg {]} \nonumber \\
&-&\sum_{q=n,p}\bigg{[}\frac{2 V_N}{3}\bigg{\{}\xi_{c,q}-\xi_{g,q}\bigg{\}}-\mu_{q}N_q\bigg{]} ;\nonumber
\end{eqnarray}

for the bulk part of the entropy:
\begin{eqnarray}
S_N^{bulk}(N_n,N_p)&=&\sum_{q=n,p}\bigg{[}\beta V_N\bigg{\{}\frac{5}{3}\bigg{(}\xi_{c,q}-\xi_{g,q}\bigg{)}+\bigg{(}U_{c,q}n_{c,q}-U_{g,q}n_{g,q}\bigg{)}\bigg{\}}-\mu_{q}N_q\bigg{]} ;\label{eq:SN}
\end{eqnarray}

for the total bulk energy of the cluster:

\begin{eqnarray}
E_N^{bulk}(N_n,N_p)
&=&V_N \left (v(n_c,\delta_c)-v(n_g,\delta_g)\right ) \label{eq:EN}\\
&+&\sum_{q=n,p}V_N  \bigg{[} \xi_{c,q}-\xi_{g,q}\bigg{]} ; \nonumber
\end{eqnarray}

and finally for the bulk part of the excitation energy:

\begin{eqnarray}
E^{*,bulk}(N_n,N_p)&=&E_N^{bulk}(N_n,N_p)-E_N^{T=0,bulk}(N_n,N_p) \\
&=&  \sum_{q=n,p}V_N  \bigg{[} \xi_{c,q}-\xi_{g,q}- \xi_{c,q}^{T=0}+ \xi_{g,q}^{T=0}\bigg{]} , \label{eq:exci}
\end{eqnarray}

where the kinetic energy density at zero temperature differs from  the standard Fermi gas expression because of the density dependent effective nucleon masses:

\begin{equation}
\xi_{ i,q}^{T=0}=\frac{3}{5}\frac{\hbar^2}{2m^*_{i,q}} n_{i,q}^{5/3}.
\end{equation}

The numerical computation of the cluster partition sum and the associated observables requires the definition of an energy functional for the nucleus interaction.
We wil adopt a recently proposed meta-modeling formulation \cite{Margueron2018} that allows reproducing different functionals and interpolating between them. For the applications shown in this paper, we will use the Sly5 empirical parameters \cite{Chabanat1998}. Detailed expressions of the effective masses $m^*_q(n,\delta)$, potential energy density $v(n,\delta)$ and the associated mean field potentials $U_q(n,\delta)=\partial v/\partial n_q$   are given in  Appendix A.
Concerning the surface free energy, this term does not enter in the gas subtraction, but it must obviously be added to have a realistic description of the cluster functional. For the present numerical applications, we use the finite nuclei extension of the meta-modelling approach \cite{Carreau2019,Carreau2019a}, detailed in Appendix B.

\subsection{Level density in the mean-field approximation}

In order to compare the effect of the thermodynamically consistent gas subtraction Eq.(\ref{eq:gas_subtraction}) to  the different phenomenological prescriptions adopted in the literature to avoid the double counting of continuum states, we have to work out the internal density of states of the nucleus. This latter is defined from the grand canonical partition sum by an inverse Laplace transform in the complex space ${\cal C}^3$. In the mean field bulk approximation we are employing, this reads:
\begin{eqnarray}
{\rho}_N(N_n,N_p,E)&=&\frac{V_N}{(2\pi i)^3}\int_{-i\infty}^{i\infty}d\beta \int_{-i\infty}^{i\infty} d\mu_n \int_{-i\infty}^{i\infty} d\mu_p \nonumber \\
&\cdot& z^{mf,N}_{\beta,\mu_n,\mu_p}
\exp \bigl [ \beta(E-\mu_n N_n -\mu_p N_p) \bigr ] \label{eq:Laplace}
\end{eqnarray}
To simplify the discussion, we will limit ourselves in this section to the case of symmetric nuclei $N_n=N_p=A/2$ and only consider the bulk part of the level density. We will also neglect Coulomb effects, such that symmetric nuclei implies $\mu_n=\mu_p=\mu$ and the dimensionality of the problem is reduced.
We note however that if the gas subtraction prescription Eq.(\ref{eq:gas_subtraction}) is used, these simplifications are not needed. Indeed to calculate the different observables Eqs.(\ref{eq:FN}),(\ref{eq:SN}), (\ref{eq:EN}), (\ref{eq:exci}), there is no need to additionally define the density of states,  that is here introduced only to compare with previous works.  Eq.(\ref{eq:gas_subtraction}) is more general, and  naturally includes isospin dependent effects in the density of states.

For convex entropies as we will be interested in, introducing the fugacity $\alpha=\beta\mu$,
the inverse Laplace transform can be calculated in the saddle point approximation \cite{BohrMottelson}  as:

\begin{eqnarray}
{\rho}_N(A,E)&=&\frac{V_N}{2\pi}\frac{ z^{mf,N}_{\beta_0,\alpha_0}
\exp (\beta_0 E-\alpha_0 A)}{|D_N|^{1/2}}\nonumber\\
&=&\frac{1}{2\pi}\frac{\exp{\Big{\{}\beta_0 \left (E-F_N^{bulk}(N_n,N_n)\right ){\Big{\}}}}}{|D_N|^{1/2}} . \label{eq:Laplace2}
\end{eqnarray}

Here, $A=N_n+N_p=A_N-n_g V_N$  is the total number of particles excluding the gas, and the only values of the intensive parameters contributing to the integrals Eq.(\ref{eq:Laplace}) are the ones verifying the ensemble equivalence conditions :

\begin{eqnarray}
-E&=&V_N \left. \frac{\partial \ln z^{mf,N}_{\beta,\alpha}   }{\partial \beta}\right |_{\alpha} \label{eq:equi1}\\
A&=&V_N\left. \frac{\partial \ln z^{mf,N}_{\beta,\alpha}   }{\partial \alpha}\right |_{\beta} . \label{eq:equi2}
\end{eqnarray}

The solutions of Eqs.(\ref{eq:equi1}),(\ref{eq:equi2}) are noted $\beta_0,\alpha_0$.
Finally, the factor $D_N$ in Eq.(\ref{eq:Laplace2} is the determinant of the $2\times 2$ susceptibility matrix calculated at $\beta=\beta_0$, $\alpha=\alpha_0$,

\begin{eqnarray}
D_N=
\begin{vmatrix}
\frac{\partial^2  \ln z^{mf,N}_{\beta,\alpha} }{\partial^2 \alpha} & \frac{\partial^2 \ln z^{mf,N}_{\beta,\alpha} }{\partial \alpha \partial \beta}  \\
\frac{\partial^2 \ln z^{mf,N}_{\beta,\alpha} }{\partial \beta \partial \alpha} & \frac{\partial^2 \ln z^{mf,N}_{\beta,\alpha} }{\partial^2 \beta}
\end{vmatrix}_{\alpha_0,\beta_0}  \ . \label{eq:determinant}
\end{eqnarray}

 Using Eqs.(\ref{eq:equi1}),(\ref{eq:equi2}), the determinant can be expressed as a function of the first derivatives of particle number and kinetic energy densities:

\begin{eqnarray}
\frac{\partial^2 \ln z^{mf,N}_{\beta,\alpha}}{\partial \alpha^2}&=&
\frac{\partial}{\partial \alpha} (n_c-n_g) \label{eq:det}\\
\frac{\partial^2 \ln z^{mf,N}_{\beta,\alpha}}{\partial \beta \partial \alpha}&=&\frac{\partial }{\partial \beta} (n_c-n_g) \nonumber\\
\frac{\partial^2 \ln z^{mf,N}_{\beta,\alpha}}{\partial \beta^2}&=& -\Big{[}\frac{2}{3}\frac{\partial }{\partial \beta}(\xi_{c}-\xi_g)+\frac{\partial}{\partial \beta}\left  (v(n_c)-v(n_g)\right )\Big{]}  , \nonumber
\end{eqnarray}

which have to be calculated at $\alpha=\alpha_0$, $\beta=\beta_0$.

These derivatives have to be calculated numerically. Indeed, for realistic energy functionals  the implicit dependence of $n$ on $\alpha$, $\beta$ due to the self-consistency, cannot be neglected with respect to the explicit dependence given by the effective chemical potential $\eta(\alpha,\beta)=\alpha -\beta U$.

The expression for the density of states Eq.(\ref{eq:Laplace2}) is still relatively involved. For this reason, it is customary to use the simpler Bethe approximation, which is obtained from the general expression Eq.(\ref{eq:Laplace2}) by neglecting all interactions leading to mean-field and effective masses, and additionally performing a low temperature Sommerfeld expansion \cite{BohrMottelson} that allows approximating the single particle level density to its value at the Fermi surface, $g(e)\approx g(e_F)$:

\begin{eqnarray}
{\rho}_{FG}(A,E^*)&=&\frac{\exp(2\sqrt{aE^*})}{4\sqrt{3}E^*} .
\label{Conventional_level_density}
\end{eqnarray}
Here, $E^*=E_N^{T>0}-E_N^{T=0}$ is the excitation energy, and the level density parameter $a$ is given by:
\begin{eqnarray}
a&=&
\frac{\pi^2}{4}\frac{A}{e_F},
\end{eqnarray}

where $e_F=\hbar^2/(2m)(3\pi^2 n/2)^{2/3}$ is the Fermi energy, and $n=n_c$ is the cluster density.
The validity of these approximations can be appreciated by comparing Eq. (\ref{Conventional_level_density}) with the total level density $\rho_{Ng}$, obtained by putting $n_g=0$ in Eqs.(\ref{eq:FN}),(\ref{eq:det}).  This we do in the next section.

Once the level density is defined through Eq.(\ref{eq:Laplace2})  for a given mass number $A$, the effect of the gas subtraction on the state counting can additionally be studied by comparing $\rho_N$ with the total level density $\rho_{Ng}$, both from  Eq.(\ref{eq:Laplace2}) with and without gas subtraction,  that we also do in the next section.

\subsection{Comparison with previous works}

\begin{figure}[!h]
\begin{center}
\includegraphics[width=0.8\textwidth,keepaspectratio=true]{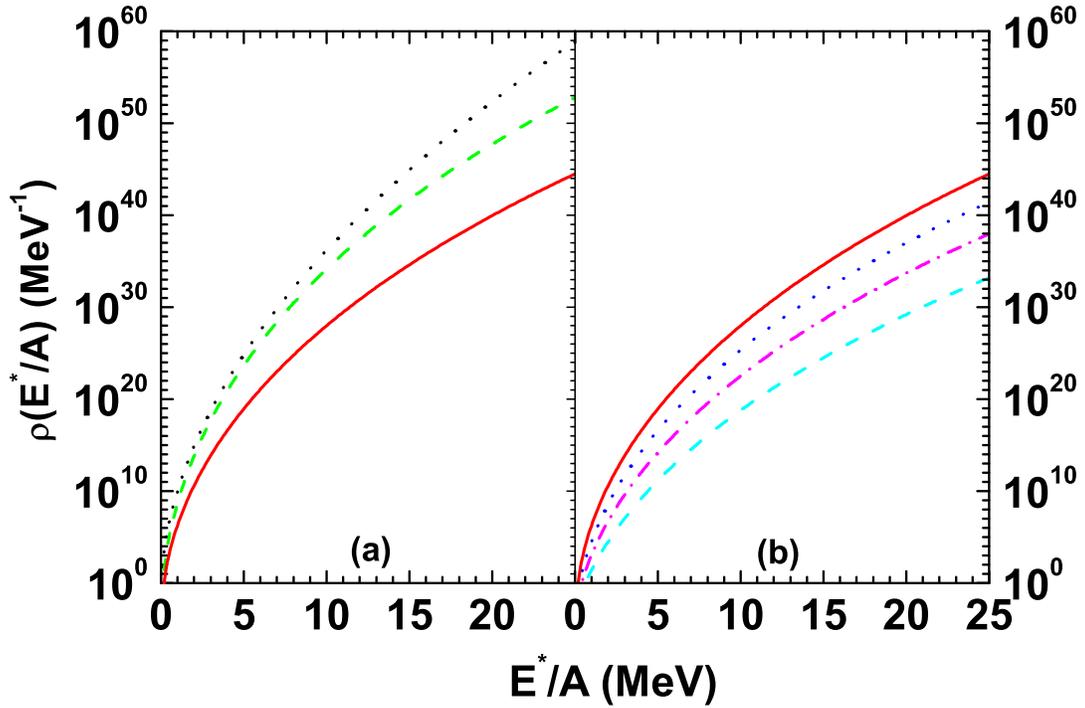}
\caption{Behavior of the bulk part of the level density for symmetric nuclei as a function of the excitation energy per baryon.\\
Left Panel: comparison of Bethe approximation Eq.(\ref{Conventional_level_density}) (black dotted line), Laplace transform Eq.(\ref{eq:Laplace0}) with  $m^*_{q}=m$=938 MeV and $U_q$=0 (green dashed line) and Laplace transform Eq.(\ref{eq:Laplace0}) with   both  $U_q$ and $m^*_{q}$ from Sly5 (red solid line).\\
Right Panel: comparison of the nucleus plus gas level density Eq.(\ref{eq:Laplace0}) (red solid line) with the continuum subtracted  Laplace transform Eq.(\ref{eq:Laplace2}), for external gas densities $n_g/n_0$=0.02 (blue dotted line), 0.05 (magenta dash-dotted line) and 0.1 (cyan dashed line). $n_0$ is the saturation density of symmetric nuclear matter. For each calculation $U_q$ and $m^*_{q}$ are included and Sly5 parameters are used.}
\label{fig:rhoe}
\end{center}
\end{figure}

In this section we compare the different prescriptions proposed in the literature to treat the internal partition sum of the clusters. Since the overcounting of continuum states only affects the bulk part of the free energy, we concentrate
on the bulk, and limit ourselves to symmetric nuclei $A=N_n+N_p=2N_n$.
The validity of the Sommerfeld expansion, and the effect of the nuclear interactions,  can be appreciated by comparing the ideal gas Bethe formula Eq.(\ref{Conventional_level_density}) obtained in the low temperature limit,  to the full inverse Laplace transform Eq.(\ref{eq:Laplace2}), calculated for the interacting nucleus plus gas system as:

\begin{equation}
{\rho}_{Ng}(A,E)=\frac{V_N}{2\pi}\frac{z^{mf}_{\beta_0,\alpha_0}
\exp (\beta_0 E-\alpha_0 A)}{|D|^{1/2}}\label{eq:Laplace0} ,
\end{equation}

where $z^{mf}_{\beta,\alpha}$ is given by Eq.(\ref{eq:zmf}) with $n_q=n_{c,q}$, and the determinant $D$ is given by the same expression Eq.(\ref{eq:determinant}) as $D^N$, with $z^{mf,N}_{\beta,\alpha}$ replaced by $z^{mf}_{\beta,\alpha}$ at the same density. As discussed above, the nuclear density $n_c$ is the solution of the self-consistency Eq.(\ref{eq:nq}), and as such in principle it depends both of the temperature and on the chemical potential, or equivalently on the gas density. However, because of the strong incompressibility of nuclear matter, if realistic functionals are used, the temperature and density dependence of $n_c$ is much weaker than the one of $n_g$, and we can  replace $n_c$ with the saturation density of nuclear matter, $n_c\approx n_0$.

To disentangle the effect of the interaction and the deviation from the Sommerfeld truncation, Eq.(\ref{eq:Laplace0}) is first computed in the same  ideal gas limit $U_q=0$, $m_q^*=m$ as the Bethe approximation  Eq.(\ref{Conventional_level_density}). The comparison between the two calculations is displayed in Fig.\ref{fig:rhoe}.
 As expected, the deviation increases with increasing excitation energy, and it is safely negligible for moderate excitations of the order $e^*\lessapprox 5$ MeV/nucleon.

The limitation of the Bethe formula clearly appears when
the nuclear interactions are accounted for in the mean-field calculation. The inclusion of an effective mass as it is required for a realistic description of nuclear structure, induces an important reduction of the level density even at low excitation energy, as we can see by comparing the full line to the dashed line in left panel of Fig.\ref{fig:rhoe}.
This important effect is fully due to the effective mass and it does not depend on the presence of a mean-field.
Indeed this latter does not affect the energy dependence of the level density, and only produces a global shift to the energy, which is exactly cancelled by the definition of the excitation energy.

Finally, the effect of the continuum subtraction can be appreciated by comparing Eq.(\ref{eq:Laplace2}) and  Eq.(\ref{eq:Laplace0})  (right panel in Fig.\ref{fig:rhoe}). Both calculations are done using the Sly5 energy functional. The importance of the gas subtraction obviously depends on the density of the external gas, and it increases with increasing gas density.

\begin{figure}[!h]
\begin{center}
\includegraphics[width=0.8\textwidth,keepaspectratio=true]{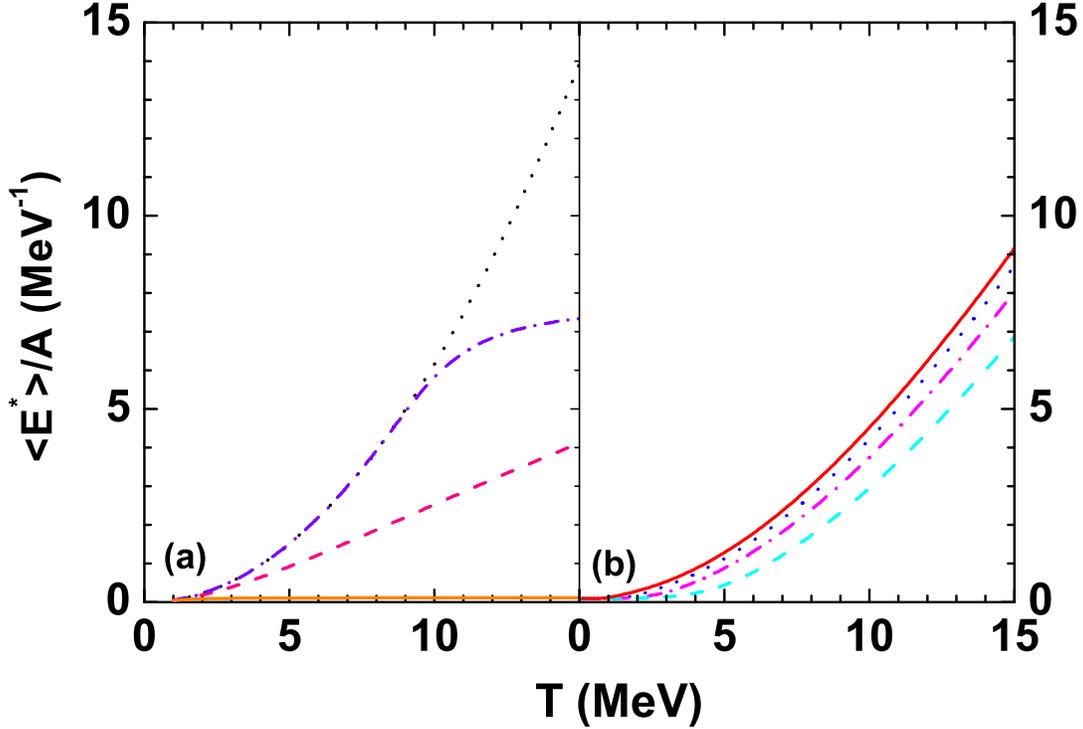}
\caption{ Average excitation energy per baryon as a function of the temperature, with different approximations for the level density.\\
Left panel: comparison of the Bethe approximation Eq.(\ref{Conventional_level_density}) (black dotted line), the Randrup and Koonin empirical correction to the Bethe approximation Eq.(\ref{eq:tlim}) with $T_{lim}=18$ MeV (pink dashed line), the Bethe approximation with $E_{cut}=S_n$ (orange solid line) and $E_{cut}=B$ (blue dash-dotted line).\\
Right panel: Continuum subtracted  Laplace transform Eq.(\ref{eq:Laplace2}), using the full Sly5 functional, for external gas densities $n_g$=0.02$n_0$ (blue dotted line), 0.05$n_0$ (magenta dash-dotted line) and 0.1$n_0$ (cyan dashed line). The red solid line represents the excitation without continuum substraction.}
\label{fig:estarT}
\end{center}
\end{figure}

We now turn to examine the consequence of the different hypotheses on physical observables. Fig.\ref{fig:estarT}  shows the average excitation energy per baryon as a function of the temperature, with different  approximations to account for the overcounting of the continuum states. In all cases, the average excitation energy of the cluster of mass $A$ at temperature $T$ reads
\begin{eqnarray}
\langle E^*(A) \rangle_T=\frac{\int_{0}^{E_{cut}}E{\rho}(A,E)\exp\Big{(}{-\frac{E}{T}}\Big{)}dE}{\int_{0}^{E_{cut}}{\rho}(A,E)\exp\Big{(}{-\frac{E}{T}}\Big{)}dE} ,
\label{Average_excitation}
\end{eqnarray}

where the different approximations consist in the different choices for the level density $\rho$, as well as in the upper limit $E_{cut}$ used for the integration. If the Sommerfeld expansion of the free Fermi gas Eq.(\ref{Conventional_level_density}) is used, and  $E_{cut}=\infty$, the well known quadratic behavior is obtained (dotted line).
A first phenomenological prescription to account for the continuum subtraction was proposed in the context of multi-fragmentation modelling \cite{Randrup1987a,Randrup1987b} employing the concept of limiting temperature, Eq.(\ref{eq:tlim}).  One can see a significant reduction in average excitation over the entire temperature range. A simpler prescription consists in keeping the Bethe expression for the level density, and introducing an upper cut for the allowed excitation energy. This cut was fixed at (or close to) the nucleon separation energy $S_n$ \cite{Fowler1978,MLB,Gross,Raduta2019}, or at the total nuclear binding energy $B$ \cite{Bondorf,Hempel2011}. The results of these prescriptions are given by the solid and dashed line in the left panel of Fig.\ref{fig:estarT}, where we have calculated $S_n$ and $B$ supposing a $^{56}Ni$ nucleus, i.e. $A=56$.
Applying an upper cut leads to a reproduction of the Sommerfeld quadratic behavior up to a temperature $T\approx E_{cut}/A$, and a saturation behavior  $\langle E^* \rangle\approx E_{cut}$ for higher temperatures. Consequently, a cut at separation energy practically amounts to ignoring the possibility of populating exciting states over the whole temperature range. Conversely, a cut at the total binding energy is fully ineffective up to temperatures of the order of 10 MeV, where the excitation energy is seen to saturate towards the total binding energy.

The effect of the approximations of the Bethe formula (Sommerfeld expansion, neglect of the effective mass) can be appreciated by comparing the dotted line of the left panel of Fig. \ref{fig:estarT} with the full Laplace transform, given by the full line of the right panel. As expected, the important reduction of the level density of the exact formula observed in Fig.\ref{fig:rhoe} leads to a global reduction of the cluster excitation energy in the whole temperature range.
Finally, the effect of the consistent gas subtraction, for different representative values of the outer gas, is also shown in the right panel of Fig. \ref{fig:estarT}. For these latter calculations, the result of Eq.(\ref{Average_excitation}) with
the continuum subtracted level density $\rho=\rho^N$ and no cut-off $E_{cut}=\infty$ coincides by construction with the grand canonical result Eq.(\ref{eq:exci}), considering  $N_n=N_p=A/2$. For practical implementations,  Eq.(\ref{eq:exci}) is clearly much simpler to calculate than Eq.(\ref{Average_excitation}) . The extra advantage of the grandcanonical formulation Eq.(\ref{eq:exci}) is that it can be extended to different neutron to proton ratios without any extra effort.

The results shown in Fig.\ref{fig:rhoe} and \ref{fig:estarT} show that the overcounting of unbound states leads to an important overestimation of the nuclear level density, which in turn produces an overestimation of the internal energy stored in fragments, already at moderate temperatures of the order of 4 MeV. None of the phenomenological recipes proposed in the literature is able to reproduce the exact state counting which is obtained when both the nuclear clusters and the nuclear gas are consistently treated within the same mean field formalism.

However, we have seen that the importance of the gas subtraction crucially depends on the relative proportion between cluster and gas. This proportion is not a free parameter, but it depends in a non-trivial way on the temperature, density and proton fraction of the system, according to the global nuclear statistical equilibrium.
This is worked out in the next section.

\section{Extension to nuclear statistical equilibrium}

\subsection{Formalism}

In the physical situations where an extended portion of diluted baryonic matter exists in thermal equilibrium at proton and neutron density $n_p, n_n$  and temperature $T$, the relative abundances of the different nuclear species  is determined by the numerical solution of the extended nuclear statistical equilibrium equations \cite{Gulminelli2015} :

\begin{eqnarray}
n_p&=&\frac{1}{2\pi^2} \Big{(}\frac{2 m^*_{g,p}}{\beta \hbar^2}\Big{)}^{3/2}F_{1/2}(\eta_{g,p}) + \sum_{N_n,N_p} N_p n_{N_n,N_p} ; \label{eq:NSE_np} \\
n_n&=&\frac{1}{2\pi^2} \Big{(}\frac{2 m^*_{g,n}}{\beta \hbar^2}\Big{)}^{3/2}F_{1/2}(\eta_{g,n})+ \sum_{N_n,N_p} N_n n_{N_n,N_p} . \label{eq:NSE_nn}
\end{eqnarray}

The first term on the r.h.s. of Eqs.(\ref{eq:NSE_np}), (\ref{eq:NSE_nn})  represent the free proton and neutron densities respectively,  see Eq.(\ref{eq:densities}). The second term gives the contribution of bound nucleons, where
$n_{N_n,N_p}$ is the number density of a cluster with $N_n$ neutrons and $N_p$ protons. We recall that these numbers do not include the contribution of nucleons in continuum states (see Eq.(\ref{eq:a_N})),
even if the unbound nucleons do contribute to the internal cluster density $n_c=A_N/V_N$ and the associated mean field, which depends on the
internal cluster density and isospin.
The independence between the free and cluster component Eqs.(\ref{eq:NSE_np}), (\ref{eq:NSE_nn}) allows simple expressions for the cluster densities  \cite{Gulminelli2015} :
\begin{equation}
n_{N_n,N_p}=  (1-u_c)\frac{g_N}{\lambda^3_N} exp\bigg{\{}-\frac{1}{T}(F_N-F_N^{id}-[T\eta_{g,p}+U_{g,p}]N_p-[T\eta_{g,n}+U_{g,n}]N_n)\bigg{\}} , \label{eq:nNZ}
\end{equation}
where the cluster free energy $F_N$ accounts for the continuum subtraction according to Eq.(\ref{eq:FN}).
In this expression,  the factor $(1-u_c)$ is an excluded volume correction that modifies the space integration associated to the center of mass free energy of each cluster:

\begin{equation}
1-u_c=\frac{V^{tot}-V_N^{tot}}{V^{tot}}=1-\frac{n_B-n_g}{\langle n_c \rangle -n_g} ,
\end{equation}

where $V^{tot}$ is the total volume occupied by the multi-component plasma, $V_N^{tot}$ is the total volume occupied by the clusters, and $\langle n_c\rangle$ is their average density:

\begin{equation}
\langle n_c \rangle =\frac{\sum_{N_n,N_p} n_{N_n,N_p} A_N}{\sum_{N_n,N_p} n_{N_n,N_p} V_N}.
\end{equation}

To get the expression Eq.(\ref{eq:nNZ}), we have additionally considered that in the full nuclear statistical equilibrium, the
translational part of the free energy is modified with respect to the expression  Eq.(\ref{eq:fideal})
\cite{Gulminelli2015,fantina2020}. Indeed, the volume $V$ appearing in Eq.(\ref{eq:fideal}) associated to the ion translational motion now corresponds to the macroscopic volume $V^{tot}$. For this reason,  the second term on the r.h.s. of Eq.(\ref{eq:fideal}) can be neglected and
we get:
\begin{equation}
F_N^{id,NSE}=-T \ln \frac{V^{tot} g_N}{\lambda^3_N},
\end{equation}
that allows the translational part of the free energy in Eq.(\ref{eq:nNZ}) to be factorized out of the exponential.

\subsection{Effect of the gas subtraction on particle abundancies}

In the calculations shown in Section \ref{sec:WScell} we have seen that the importance of the continuum subtraction obviously depends on the value of the external gas density, since the double counting concerns the single particle gas states.
In realistic physical situation, the gas density is not a free parameter but it depends on the global temperature, pressure (or equivalently global density) and proton fraction of the system.

\begin{figure}[b]
\begin{center}
\includegraphics[width=0.7\textwidth,keepaspectratio=true]{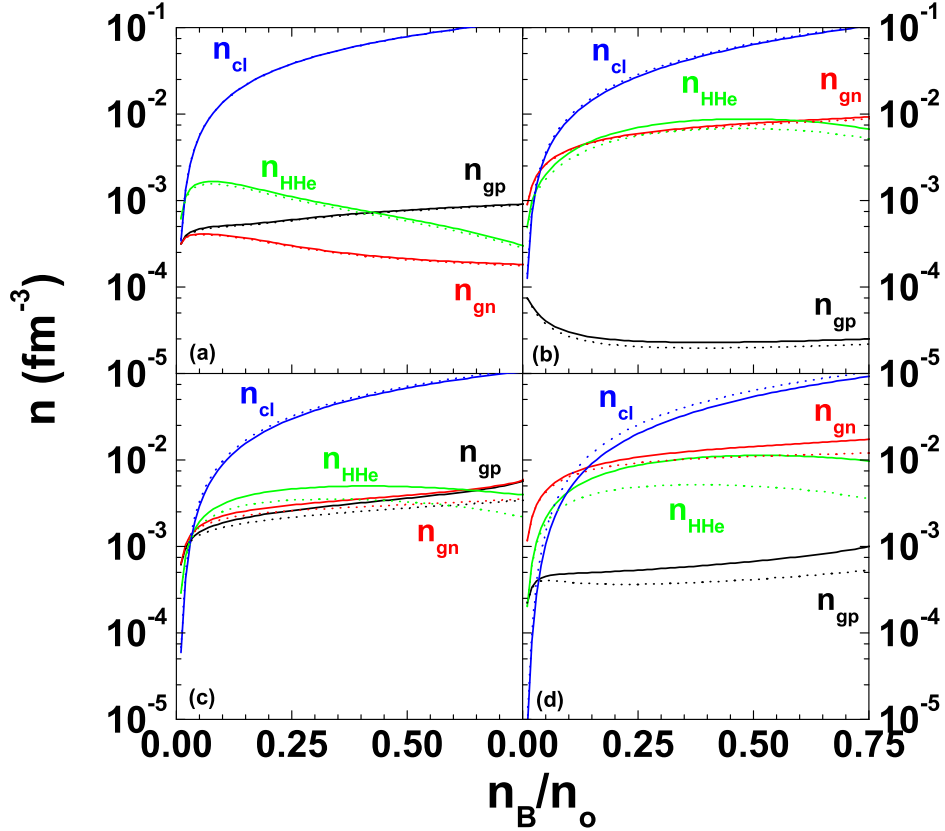}
\caption{Free neutron $n_{gn}$ (black lines), proton $n_{gp}$ (red lines), bound nucleon density in $H$ and $He$ isotopes $n_{HHe}$ (green lines) and in heavier elements $n_{cl}$ (blue lines) as a function of the global density, as predicted in full thermodynamic equilibrium at two representative temperatures $T=5$ MeV (upper panels) and $T=10$ MeV (lower panels), and two representative proton fraction $Y_p=0.5$ (left panels) and $Y_p=0.2$ (right panels). Dotted (full) lines give predictions without (with) the consistent continuum subtraction. }
\label{fig:ngas_ncluster}
\end{center}
\end{figure}

The evolution of the external gas density as a function of the thermodynamic conditions of the statistical equilibrium can be obtained by numerically solving the coupled equations Eqs.(\ref{eq:NSE_np}), (\ref{eq:NSE_nn}). The resulting behavior is reported in Fig.\ref{fig:ngas_ncluster} for different temperatures, densites and proton fraction representative of the typical conditions that can be encountered in heavy ion collisions (left panels) and in supernova and proto-neutron star matter (right panels). In this figure, the density of nucleons bound in the different $H$ and $He$ isotopes is defined as $n_{HHe}= \sum_{N_n>0}\left ( (N_n+1) n_{N_n,1}+ (N_n+2) n_{N_n,2}\right ) $, and the cluster density is  $n_{cl}=n_B-n_g-n_{HHe}$.
 We can see that, for a given baryonic density and proton fraction, proton and neutron densities as well as $n_{HHe}$ increase with temperature, therefore the cluster density decreases. For symmetric matter  (left panels),
the density of unbound nucleons is several order of magnitude lower than the density of clusters except at the lowest densities, and we expect that the consistent continuum subtraction will only be important below $\approx n_0/10$. For neutron rich matter beyond the neutron driplines (right panels), the gas contribution is typically dominant over the clusterized one, and we expect more important effects of the correct state counting. We can also observe that at those low proton fraction the free proton contribution is systematically negligible even at very high temperature, the continuum being essentially represented by neutron excitations.
As we can see from the comparison between full and dotted lines, these general features are very robust and show a very limited dependence on the finite temperature state counting.

\begin{figure}[hbtp]
\begin{center}
\includegraphics[width=0.7\textwidth,keepaspectratio=true]{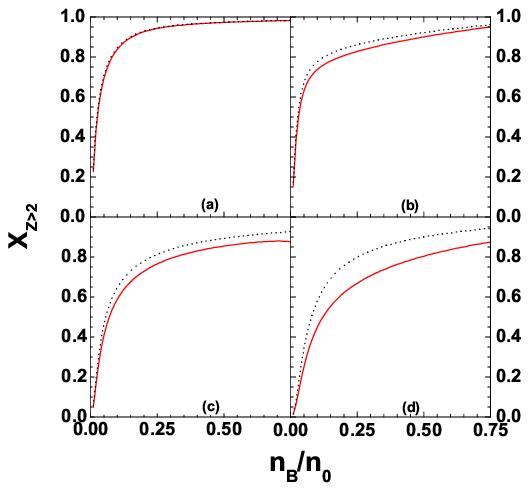}
\caption { Mass fraction of elements heavier than Helium as a function of the global density, as predicted in full thermodynamic equilibrium at two representative temperatures $T=5$ MeV (upper panels) and $T=10$ MeV (lower panels), and two representative proton fraction $Y_p=0.5$ (left panels) and $Y_p=0.2$ (right panels). Black dotted ( red full) lines give predictions without (with) the consistent continuum subtraction.  }
\label{fig:xheavy}
\end{center}
\end{figure}

\begin{figure}[hbtp]
\begin{center}
\includegraphics[width=0.7\textwidth,keepaspectratio=true]{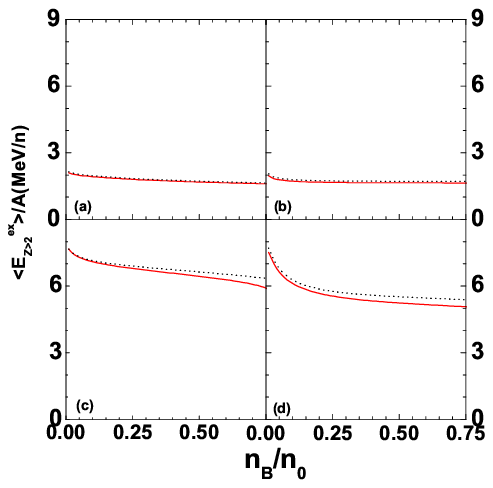}
\caption{ Average excitation energy of elements heavier than Helium as a function of the global density, as predicted in full thermodynamic equilibrium at two representative temperatures $T=5$ MeV (upper panels) and $T=10$ MeV (lower panels), and two representative proton fraction $Y_p=0.5$ (left panels) and $Y_p=0.2$ (right panels). Black dotted ( red full) lines give predictions without (with) the consistent continuum subtraction. }
\label{fig:estar}
\end{center}
\end{figure}

The effect of the gas subtraction on the observables can be appreciated  more clearly from Figs.\ref{fig:xheavy} and \ref{fig:estar}, which report, in the same thermodynamic conditions as in Figure \ref{fig:ngas_ncluster}, the mass fraction bound in clusters and their average excitation energy per nucleon of the clusters.

In symmetric matter at moderate temperatures, the double counting of continuum states is fully negligible at all densities, and almost identical  predictions are obtained if the gas partition sum is subtracted from the cluster level counting or not. However this is not the case any more if the temperature is higher, or the system is strongly asymmetric: at the highest temperature and lowest proton fraction, such as they can be encountered in the early evolution of the proto-neutron star, the cluster mass fraction drops from $\approx 60\%$ to $\approx 20\%$ at the highest density close to the crust-core transition, if the gas subtraction is taken into account. Figure \ref{fig:estar} additionally shows that the energy stored in internal cluster degrees of freedom is systematically overestimated if the effect of the gas is not accounted for. As expected, the effect increases with increasing temperature and decreasing proton fraction.

The double counting of continuum states between the gas and cluster component does not directly affect the light particles with $N_p=1$ (Hydrogen) and $N_p=2$ (Helium), because such light elements  have a sparse spectrum with very few or no excited states. In our model, we do not consider any excited state for such clusters, that is we consider $F_N=F_N^{id,NSE}-B(N_n,N_p)$, where $B$ is the experimental binding energy of the considered Helium or Hydrogen isotope. However, because of the mass and charge conservation laws Eqs.(\ref{eq:NSE_np}), (\ref{eq:NSE_nn}), the continuum subtraction may indirectly affect the abundancy of those light cluster, which as we saw in Fig.\ref{fig:ngas_ncluster} dominate the composition of very hot matter except at the highest densities.

\begin{figure}[hbtp]
\begin{center}
\includegraphics[width=0.7\textwidth,keepaspectratio=true]{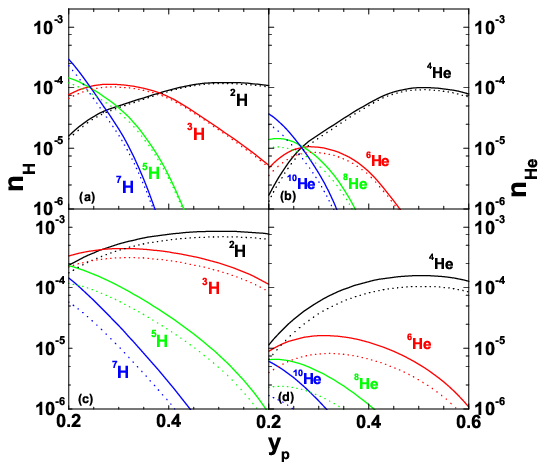}
\caption{Density of Hydrogen (left panel) and Helium (right panel) isotopes  as a function of the global proton fraction, as predicted in full thermodynamic equilibrium  for $n_B=0.3 n_0$ and  $T=5$ MeV (upper panels) and  $T=10$ MeV (lower panels). Dotted (full) lines give predictions without (with) the consistent continuum subtraction.  }
\label{fig:light}
\end{center}
\end{figure}

This is shown in Fig.\ref{fig:light}, which displays as a function of the global proton fraction the density of the different $H$ and $He$ isotopes
in hot and dense matter, for $n_B=0.3 n_0$ and  two different temperatures. Concerning Hydrogen (right panel), we can see that deuterons dominate over heavier isotopes for relatively symmetric matter, but tritons take over for asymmetric matter and heavier resonances dominate in extremely neutron rich matter. Those very loosely bound resonances disappear at the highest temperatures, but deuterons and tritons survive with a non negligible fraction at temperatures as high as 10 MeV. The absolute fraction of $He$ is an order of magnitude lower than the one of $H$, but concerning the relative abundancy of the different isotopes a similar trend can be observed. Specifically, $^4He$ dominates in symmetric matter and for moderate temperatures, while more neutron rich isotopes tend to prevail if the global proton fraction decreases. Interestingly, loosely bound neutron rich resonances dominate the Hydrogen and Helium composition of very rich neutron matter if the temperature is not too high. This feature is not accounted in the most popular EoS models for astrophysical applications \cite{LS,Shen,Shen2011}. Concerning the effect of the continuum subtraction, we can see that the suppression of the high energy continuum states of the heavy clusters naturally leads to an increased importance of the $H$ and $He$ isotopes, for which no correction was applied. This is particularly clear at high temperature and low proton fraction, where the continuum subtraction procedure is most effective.
This effect is however globally very small whatever the thermodynamic condition.

These results are in good qualitative agreement with previous works \cite{Hempel2012,Raduta2019}, but they should be taken with care, because in the calculation we have considered experimental vacuum binding energies for all light clusters, while it is known from both theoretical \cite{Typel2010,Ropke2011,Pais2019} and experimental works \cite{Qin2012,Hempel2015,Bougault2020,Pais2020} that the effect of the occupied gas orbitals is an in-medium modification of the ground state binding energy of those clusters, effect that cannot be treated with the present mean-field approach. In particular, the use of the Beth-Uhlenbeck formula and the corresponding virial expressions to treat the contributions of the continuum seems to give larger deviations, see refs.\cite{Yudin,Ropke}.

\section{Conclusions}

In this paper, we have proposed
a well-defined protocol to consistently treat the internal nuclear degrees of freedom in the finite temperature sub-saturation equation of state, that is needed to model different dynamical processes such as supernova collapse, proto-neutron star cooling, neutron star mergers, and heavy ion collisions.
Specifically, we have worked out a simple analytical expression for the internal nuclear partition sum in the presence of continuum states, which is based on a decomposition of the nuclear energy into a bulk and a surface part in the spirit of the compressible liquid drop model.

Indeed, the relative proportion between free nucleons and nucleons bound in clusters in dense and hot nuclear matter is governed by the
 nuclear partition sum; at high temperature, this latter includes unbound continuum states that should not be double counted with the states of the free nucleon gas. We have shown that an exact subtraction of the continuum contribution can be performed if the problem is treated in the self-consistent mean-field approximation.  The resulting nuclear level density obtained by an inverse Laplace transform in the saddle point approximation was compared to different phenomenological prescriptions proposed in the equation of state literature, and we have observed that none of them correctly reproduces the exact gas subtraction at high temperature.

In a second part of the paper, we have worked out the effect of gas subtraction on the composition of stellar matter in different thermodynamic conditions, and shown that it leads to an increased fraction of unbound nucleons and light clusters, with respect to calculations that do not account for this effect.

This method of consistently treating the internal nuclear degrees of freedom in the thermodynamical equilibrium reduces the model dependence of the sub-saturation equation of state to the choice of the nuclear energy functional, which is still affected by important uncertainties as far as isospin asymmetric matter is concerned.
The calculations presented in this paper were performed using the Sly5 functional, but the expression can be used for any non-relativistic or relativistic energy functional, provided the empirical parameters of the equation of state for homogeneous matter are known. The study of such possible model dependence induced by the nuclear energy functional is left for future work.

The main limitation of the present formalism is that the effect of the nuclear gas can only be computed on heavy nuclei that can be realistically described by the mean-field approach. Concerning  light clusters such as Hydrogen and Helium isotopes, the vacuum free energy is assumed and the in-medium modifications induced by the nucleon gas is simply treated in the excluded volume approximation. In the future, we plan to fix the correction to the light cluster free energy using recent constraints from heavy ion collisions \cite{Pais2020}.
We expect that a modified composition of stellar matter, due to the inclusion of gas corrections to both heavy and light clusters, might have consequences on neutrino transport properties in hot and dense matter, that is the final aim of this work.

\section{Acknowledgements}
We acknowledge the support of “IFCPAR/CEFIPRA” Project No. 5804-3. Valuable discussions with G. Chaudhuri of VECC are highly appreciated. We warmly thank T. Carreau for providing the surface free energy parameters  for the Sly5 functional.

\section{Appendix}

In the following appendixes, the detailed expressions of the nuclear energy functional used for the presented numerical applications are given.

\subsection{Meta-modelling of bulk matter}

The calculation of the bulk free energy Eq.(\ref{eq:FN}), total energy Eq.(\ref{eq:EN}), and excitation energy Eq.(\ref{eq:exci}) requires the specification of the mean field potentials $U_{i,q}$ and energies $v(n_i,\delta_i)$.
In the meta-modelling framework \cite{Margueron2018}, the expression of potential energy per particle that can be adapted to different effective interactions and energy functionals  is given by:
\begin{eqnarray}
v(n,\delta)&=&\sum_{k=0}^{N}\frac{1}{k!}(v^{is}_{k}+v^{iv}_{k}\delta^2)x^{k}\nonumber\\
&+&(a^{is}+a^{iv}\delta^2)x^{N+1}\exp(-b\frac{n}{n_0}) ,
\label{ELFc_potential}
\end{eqnarray}
where $x=(n-n_0)/3n_0$ and  $n_0$ is the saturation density of symmetric nuclear matter.
For this paper, we choose $N=4$ and $b=10ln2$. This value of $b$ leads to a good reproduction of the Sly5 functional which is used for the numerical applications presented in this paper. The model parameters $v_k^{is(iv)}$ can be linked with a one-to-one correspondence to the usual EoS empirical parameters, via:
\begin{eqnarray}
v^{is}_{0}&=&E_{sat}-t_0(1+\kappa_0)\nonumber\\
v^{is}_{1}&=&-t_0(2+5\kappa_0)\nonumber\\
v^{is}_{2}&=&K_{sat}-2t_0(-1+5\kappa_0)\nonumber\\
v^{is}_{3}&=&Q_{sat}-2t_0(4-5\kappa_0)\nonumber\\
v^{is}_{4}&=&Z_{sat}-8t_0(-7+5\kappa_0)
\label{Isoscalar_parameters}
\end{eqnarray}
\begin{eqnarray}
v^{iv}_{0}&=&E_{sym}-\frac{5}{9}t_0[(1+(\kappa_0+3\kappa_{sym})]\nonumber\\
v^{iv}_{1}&=&L_{sym}-\frac{5}{9}t_0[(2+5(\kappa_0+3\kappa_{sym})]\nonumber\\
v^{iv}_{2}&=&K_{sym}-\frac{10}{9}t_0[(-1+5(\kappa_0+3\kappa_{sym})]\nonumber\\
v^{iv}_{3}&=&Q_{sym}-\frac{10}{9}t_0[(4-5(\kappa_0+3\kappa_{sym})]\nonumber\\
v^{iv}_{4}&=&Z_{sym}-\frac{40}{9}t_0[(-7+5(\kappa_0+3\kappa_{sym})] \ ,
\label{Isovector_parameters}
\end{eqnarray}
where  $E_{sat}$, $K_{sat}$, $Q_{sat}$ and $Z_{sat}$ are saturation energy, incompressibility modulus,  isospin symmetric skewness and  kurtosis respectively and $E_{sym}$, $L_{sym}$, $K_{sym}$, $Q_{sym}$ and $Z_{sym}$ are symmetry energy, slope, and associated incompressibility,  skewness and  kurtosis respectively.
Concerning the $\kappa_0$ and $\kappa_{sym}$, they govern the density dependence of the neutron and proton effective mass according to:
\begin{equation}
\frac{m_q}{m^*_{q}(n,\delta)}=1+(\kappa_0 \pm \kappa_{sym}\delta)\frac{n}{n_{0}},
\end{equation}
with $q=n,p$.
For the applications presented in this paper, all the parameters are taken from the Sly5 functional.\\

The baryonic density of cluster with isospin asymmetry $\delta_c=\frac{N_n-N_p}{N_n+N_p}$ is approximated to the correspoding saturation density at finite asymmetry according to:
\begin{eqnarray}
n_c(\delta)=n_0\bigg{(}1-\frac{3L_{sym}\delta_c^2}{K_{sat}+K_{sym}\delta_c^2}\bigg{)}.
\end{eqnarray}

\subsection{ Surface free energy}

For the surface free energy, we use the prescription proposed in ref.\cite{LS,Carreau2019,Carreau2019a} on the basis of Thomas-Fermi calculations with extreme isospin ratios:
\begin{eqnarray}
F_N^{surf}&=&4\pi r^2_c A_N^{2/3}\sigma(y_{c,p},T)
\label{Surface}
\end{eqnarray}
with $A_N=N_n+N_p+n_g V_N$,  $r_c={\{} 3/(4\pi n_c){\}}^{1/3}$, $y_{c,p}=N_p/(N_p+N_n)$ and
\begin{eqnarray}
\sigma(y_{c,p},T)=\sigma_0 h\bigg{(}\frac{T}{T_c(y_{c,p})}\bigg{)}\frac{2^{p+1}+b_s}{y_{c,p}^{-p}+b_s+(1-y_{c,p})^{-p}}
\end{eqnarray}
where $\sigma_0$ represents the surface tension of symmetric nuclear matter and $b_s$ and $p$  govern the isospin dependence. For the Sly5 functional, the parameters were optimized in \cite{Carreau2019} as $\sigma_0=1.09191$, $b_s=15.36563$ and $p=3.0$. The temperature dependence is incorporated by
\begin{eqnarray}
h\bigg{(}\frac{T}{T_c(y_{c,p})}\bigg{)} &=& \bigg{[}1-\bigg{(}\frac{T}{T_c(y_{c,p})}\bigg{)}^2\bigg{]}^2
\hspace{1.2cm}\mbox{for}\hspace{0.3cm} T \le T_c(y_{c,p})\nonumber\\
&=& 0 \hspace{4.1cm}\mbox{for}\hspace{0.3cm} T > T_c(y_{c,p}).
  \end{eqnarray}
$T_c(y_{c,p})$ is the  critical temperature of the nuclear liquid-gas phase transition approximated as (in MeV units) \cite{LS}:
\begin{eqnarray}
T_c(y_{c,p})=87.76\bigg{(}\frac{K_{sat}}{375}\bigg{)}^{1/2}\bigg{(}\frac{0.155}{n_0}\bigg{)}^{1/3}y_{c,p}(1-y_{c,p}) ,
\end{eqnarray}
where $K_{sat}$, $n_0$ are expressed in MeV and fm$^{-3}$ respectively.



\end{document}